\documentclass[review,authoryear]{elsarticle}

\usepackage{graphicx}
\usepackage[tight,footnotesize, nooneline]{subfigure}

\usepackage{adjustbox} 
\usepackage[hidelinks]{hyperref} 

\usepackage{amssymb}
\usepackage{amsthm}
\usepackage{amsmath}

\usepackage{lineno}
 
\usepackage{multirow}

\usepackage[flushleft]{threeparttable} 

\journal{arXiv}

\newcommand\copyrighttext{%
   \textcopyright\ 2018 IEEE.}

\begin{document}

\begin{frontmatter}



\title{The effect of tissue physiological variability on transurethral ultrasound therapy of the prostate}


\author[Affil1]{Visa Suomi \corref{cor1}}
\author[Affil2]{Bradley Treeby}
\author[Affil3]{Jiri Jaros}
\author[Affil4]{Jani Saunavaara}
\author[Affil1]{Aida Kiviniemi}
\author[Affil1]{Roberto Blanco}

\address[Affil1]{Department of Radiology, Turku University Hospital, Kiinamyllynkatu 4-8, 20521 Turku, Finland}
\address[Affil2]{Department of Medical Physics and Biomedical Engineering, University College London, Gower Street, London, WC1E 6BT, UK}
\address[Affil3]{Centre of Excellence IT4Innovation, Faculty of Information Technology, Brno University of Technology, Brno, Czech Republic}
\address[Affil4]{Department of Medical Physics, Turku University Hospital, Kiinamyllynkatu 4-8, 20521 Turku, Finland}
\cortext[cor1]{Corresponding Author: Visa Suomi, Department of Radiology, Turku University Hospital, Kiinamyllynkatu 4-8, 20521 Turku, Finland; Email, visa.suomi@tyks.fi}

\begin{abstract}
Therapeutic ultrasound is an investigational modality which could potentially be used for minimally invasive treatment of prostate cancer. Computational simulations were used to study the effect of natural physiological variations in tissue parameters on the efficacy of therapeutic ultrasound treatment in the prostate. The simulations were conducted on a clinical ultrasound therapy system using patient computed tomography (CT) data. The values of attenuation, perfusion, specific heat capacity and thermal conductivity were changed within their biological ranges to determine their effect on peak temperature and thermal dose volume. Increased attenuation was found to have the biggest effect on peak temperature with a 6.9\% rise. The smallest effect was seen with perfusion with $\pm$0.2\% variation in peak temperature. Thermal dose was mostly affected by specific heat capacity which showed a 20.7\% increase in volume with reduced heat capacity. Thermal conductivity had the smallest effect on thermal dose with up to 2.1\% increase in the volume with reduced thermal conductivity. These results can be used to estimate the interpatient variation during the therapeutic ultrasound treatment of the prostate.
\end{abstract}


\end{frontmatter}

\copyrighttext

\pagebreak



\section*{Introduction}

Prostate cancer is the second most common cancer occurring in men, with an estimated 1.1 million people diagnosed worldwide in 2012 \citep{stewart2014world}. In the same year, approximately 0.3 million people died due to the disease, making prostate cancer the fifth most common cause of cancer death among men. To put these figures into perspective, prostate cancer accounts for approximately 15\% of all cancer incidences and 7\% of all cancer related deaths in men \citep{stewart2014world}. Patients typically experience symptoms, such as urinary problems and pelvic pain, which reduce their quality of life. Therefore, early diagnosis and effective treatment of the disease are essential for the well-being and survival of the patients. 

Therapeutic ultrasound is a treatment modality which could potentially provide minimally invasive therapy for prostate cancer patients. The treatment can be delivered through a transurethral route \citep{burtnyk2015magnetic, ramsay2017evaluation} whereby the ultrasound probe is inserted into the prostate through the urethra. The benefit of this technique is that the heat can be delivered directly to the target location without any intervening tissue layers in between. The therapy can then be delivered to either the complete prostate or parts of it by controlling the rotation of the ultrasound probe.

The initial clinical evidence from transurethral ultrasound therapy of the prostate has shown variability in efficacy \citep{chin2016magnetic, ramsay2017evaluation}, which can be attributed to several factors. One possible reason might be the physiological differences between patients. There exists natural variation in the acoustic and thermal properties of the prostate \citep{parker1993elastic, van2002prostate, patch2011specific, hasgall2015database}, which might affect the treatment efficacy. It has been shown that these parameters have an effect on heating and lesion creation in other therapeutic ultrasound treatments \citep{billard1990effects, chen1993effect, burtnyk2010simulation}.

The aim of this research is therefore to find out how much the efficacy of ultrasound therapy in the prostate is affected by the natural variation in the acoustic and thermal properties of the prostate. This is done by conducting nonlinear ultrasound and thermal simulations on a clinical patient image data by varying the values of attenuation, perfusion, specific heat capacity and thermal conductivity of the prostate within their physiological ranges. The results help to understand the scale of interpatient variability that can be expected to occur during clinical treatments.

\section*{Therapeutic ultrasound simulations}

\subsection*{Acoustic and thermal simulation models}

\begin{figure*}[b!]
    \centering
    \subfigure[]
    {
        \includegraphics[height=2.9cm]{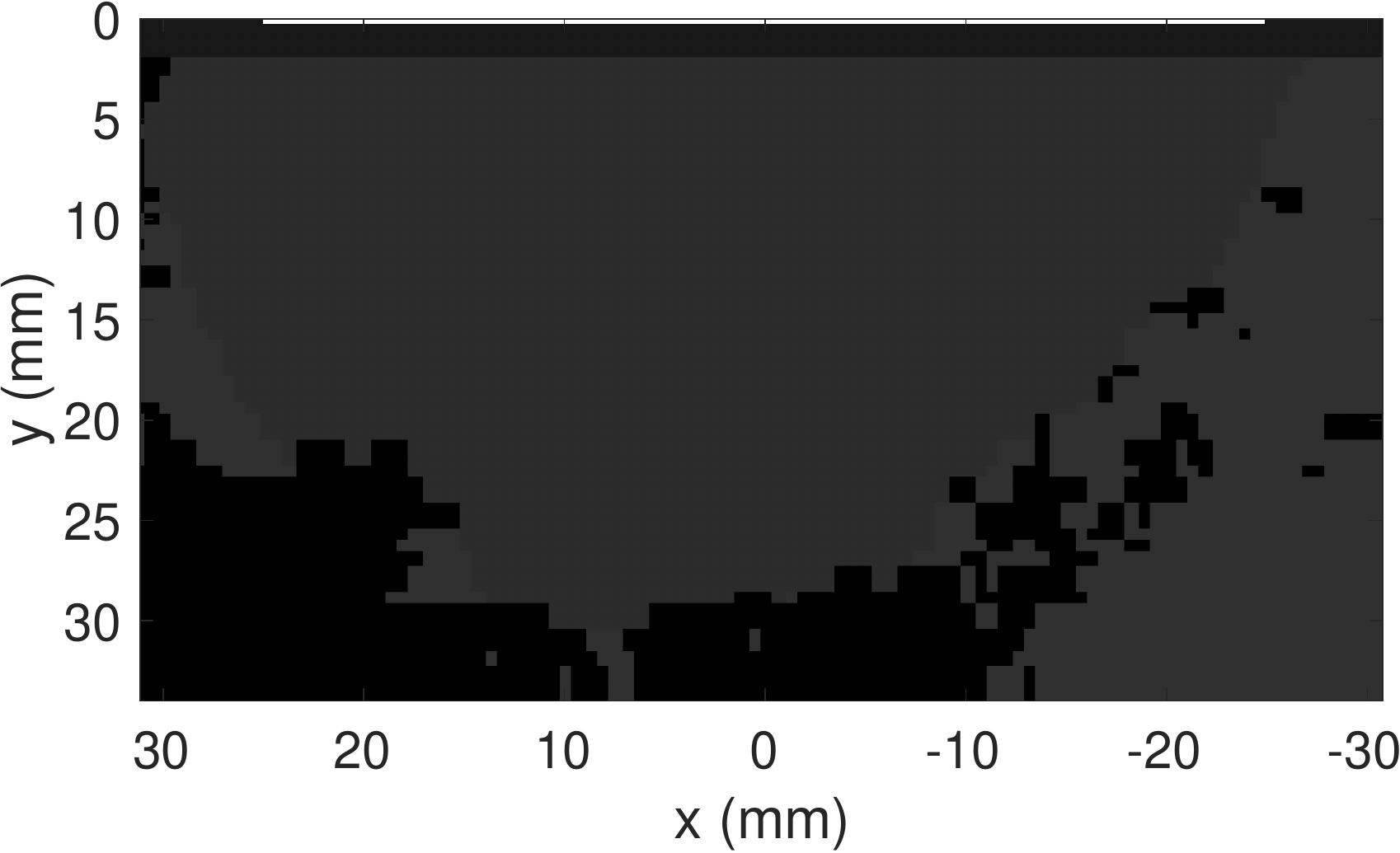}
    }
    \subfigure[]
    {
        \includegraphics[height=2.9cm]{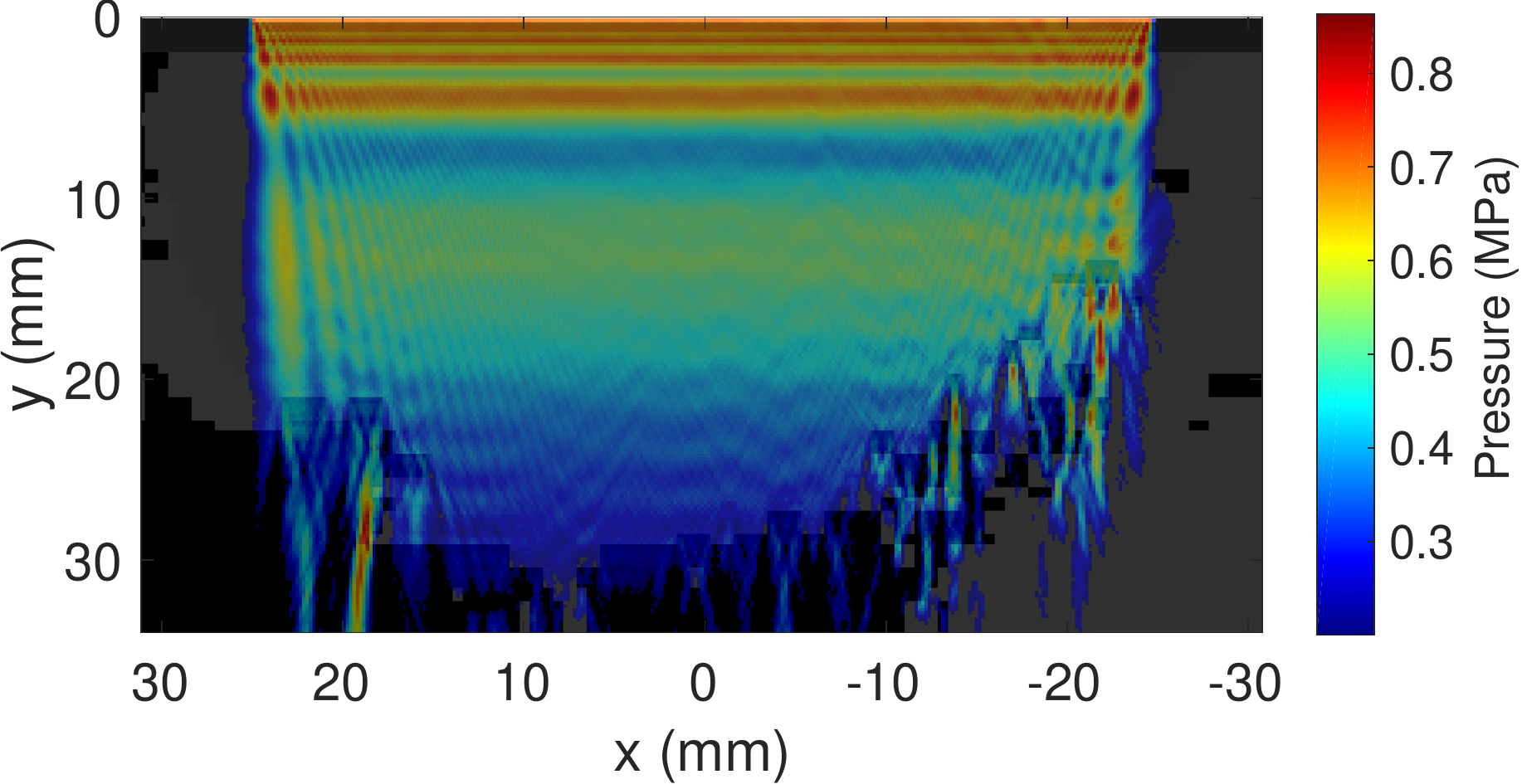}
    }
    \subfigure[]
    {
        \includegraphics[height=2.9cm]{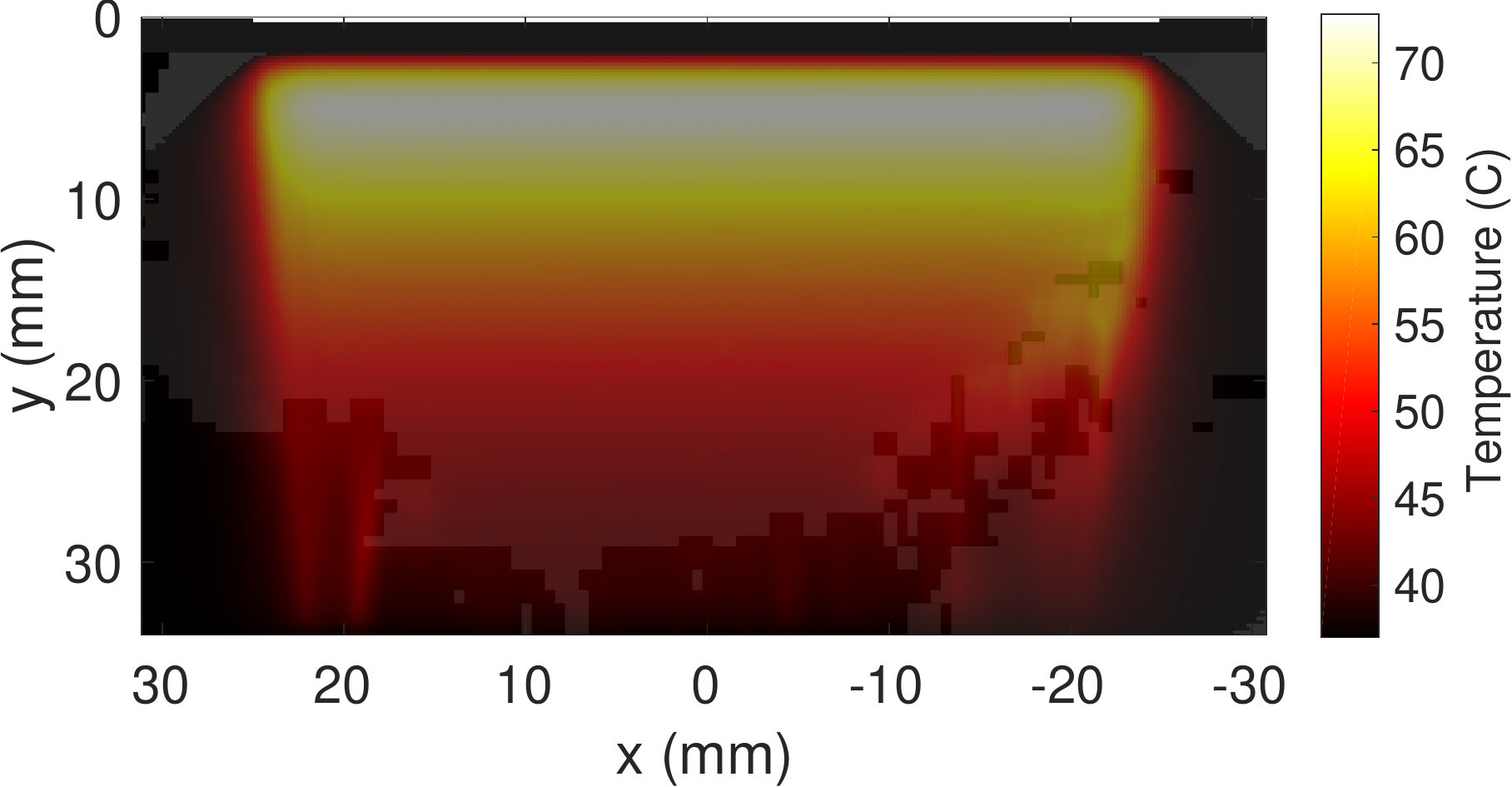}
    }
    \caption{Segmented computed tomography (CT) slice of half of the prostate (grey). The tissue areas surrounding the prostate were segmented as fat (black) and muscle (light grey). The ultrasound probe was positioned along the urethra in the middle of the prostate (the white area on top of the image is the transducer). (b) Simulated pressure and (c) temperature fields in the prostate during a 20-second sonication.}
    \label{fig:planes}
\end{figure*}

The simulation geometry was derived using a three-dimensional computed tomography (CT) dataset of a clinical patient treated at the Turku University Hospital, Finland. The ethical permission for the study (ETMK: 152/1801/2016) was obtained from the Ethics Committee of Hospital District of Southwest Finland. Intensity thresholds were first used to automatically segment the CT data into fat, bone and muscle tissue after which the prostate was segmented manually.

The therapeutic ultrasound probe was modelled on a clinical Tulsa-Pro system (Profound Medical, Mississauga, Canada) \citep{burtnyk2015magnetic}. The system has 10 rectangular unfocused transducer elements, which are located inside the ultrasound probe. The element size is 4.5 mm $\times$ 5.0 mm with 0 mm element spacing which results in a total transmitting surface area of 4.5 mm $\times$ 50.0 mm. The diameter of the ultrasound probe is approximately 5 mm with the transducer being 2 mm inside the probe. The transducer was operated at 4 MHz frequency with continuous wave transmission. The ultrasound probe was positioned along the urethra in the middle of the prostate so that all transducer elements were inside the prostate. A visualisation of the prostate together with the inserted ultrasound probe is presented in Figure \ref{fig:planes}(a) where the craniocaudal direction is along the positive x-axis.

The ultrasound simulations were conducted using a parallelised version of the open source k-Wave Toolbox \citep{treeby2012modeling, jaros2016full}. The code solves a set of coupled first-order partial difference equations based on the conservation laws and a phenomenological loss term that accounts for acoustic absorption with a frequency power law. The governing equations are equivalent to a generalised version of the Westervelt equation that accounts for second-order acoustic nonlinearity, power law acoustic absorption, and a heterogeneous distribution of material properties (sound speed, density, nonlinearity and absorption coefficient).

The thermal simulations were conducted by solving the Pennes bioheat transfer equation. The solution took into account the specific heat capacity, thermal conductivity and the perfusion in different tissues. The heating rate was calculated using the harmonic components of the nonlinear ultrasound field. This was done in order to accurately replicate the increased heating effect in the focal area of the ultrasound field due to nonlinearity.

\subsection*{Simulation parameters and execution}

\begin{table}[t!]
  \centering
  \caption{Acoustic simulation parameters}
    \begin{tabular}{lcccc}
    \hline
          		& Density   		& Sound speed   	& Attenuation 				& B/A 	\\
          		& (kg/m$^{3}$) 	& (m/s) 			& (dB/MHz$^{1.1}$/cm)		& 		\\
    \hline
    Prostate		& 1045  			& 1561			& 0.78 $\pm$ 0.24   			& 7.5 	\\
    Muscle	 	& 1050  			& 1547  			& 1.09   					& 7.2 	\\
    Fat   		& 950   			& 1478  			& 0.48  						& 10.1 	\\
	Water 		& 1000  			& 1520  			& 0.00217 					& 5.2 	\\
    \hline
    \end{tabular}
  \label{tab:acoustic_parameters}
\end{table}

The ultrasound simulations were run on a computing cluster at CSC - IT Centre for Science, Finland, using 256 cores, 90 GB memory and approximately 3 hours per simulation. The size of the computational grid was 1280 $\times$ 1536 $\times$ 256 grid points, i.e., 5.9~cm $\times$ 7.1~cm $\times$ 1.2~cm, which supported harmonic frequencies up to 16 MHz (i.e., four harmonics with the sonication frequency of 4 MHz). Temporal resolution was set to 30 points per wavelength which corresponded to a time step of 8.3 ns.

In total, three different acoustic simulations were run using the tissue parameters in Table \ref{tab:acoustic_parameters} \citep{mast2000empirical, hasgall2015database, parker1993elastic}. In addition to the `baseline' simulation, which was used as a reference with mean values, the attenuation of the prostate was varied by $\pm$0.24~dB/MHz/cm which corresponds to one standard deviation (SD) variation in the prostate tissue \citep{parker1993elastic}.

The thermal simulations were run in Matlab R2017a (MathWorks, Natick, Massachusetts, United States) on a local desktop computer. The grid resolution was decimated by a factor of 4 for computational efficiency and a time step of 0.25~s was used. The thermal simulations were conducted using the tissue parameters in Table \ref{tab:thermal_parameters} \citep{van2002prostate, patch2011specific, hasgall2015database}. Each simulation was run using 20-second heating time followed by 40 seconds of cooling. In addition to the baseline simulation with mean values, the thermal conductivity, specific heat capacity and perfusion rate of the prostate were varied by $\pm$0.03~W/m/K, $\pm$300~J/kg/K and $\pm$1.3~kg/m$^{3}$/s, respectively, which correspond to one SD change measured in the prostate \citep{van2002prostate, patch2011specific, hasgall2015database}. The temperature of the water inside the ultrasound probe was held constant at 21~$^{\circ}$C to mimic the cooling effect of the room temperature water flowing through the clinical system. The perfusion rate was set to zero for tissue regions which reached a thermal dose of 240 cumulative equivalent minutes at 43~$^{\circ}$C (CEM).

\begin{table}[t!]
  \centering
  \caption{Thermal simulation parameters}
    \begin{tabular}{lccc}
    \hline
          			& Thermal			& Specific		& Perfusion 			\\
          			& conductivity		& heat capacity	& rate				\\
          			& (W/m/K) 			& (J/kg/K) 		& (kg/m$^{3}$/s)		\\
    \hline
    Prostate			& 0.51 $\pm$ 0.03  	& 3400 $\pm$ 300	& 1.7 $\pm$ 1.3		\\
    Muscle			& 0.49 				& 3421  			& 0.6				\\
    Fat				& 0.21				& 2348			& 0.6				\\
    Water			& 0.60				& 4178			& 0					\\
    Blood 			& N/A      			& 3617  			& N/A 				\\
    \hline
    \end{tabular}
  \label{tab:thermal_parameters}
\end{table}

\section*{Results}

A visualisation of the segmented CT data together with the simulated ultrasound and temperature fields are shown in Figure \ref{fig:planes}. The ultrasound field in Figure \ref{fig:planes}(b) can be seen to exit the transducer into the prostate with the highest pressure region occurring near the transducer face. Some high pressure regions can also be seen in the regions where the ultrasound field is transmitted from the prostate tissue into the fat and muscle. Similarly, the temperature field in Figure \ref{fig:planes}(c) can be seen forming close to the transducer where the high pressure regions are located. The peak temperatures were observed to occur approximately at y = 5 mm from the transducer face. The locations at y = 2 mm and closer are at 21~$^{\circ}$C due to the water cooling.

\begin{table*}[t!]
  \centering
  \caption{The variation in the peak temperatures and thermal dose volumes}
  \adjustbox{max width=\textwidth}{
    \begin{tabular}{llrrrrrrrrr}
    \hline
          &		& \multicolumn{1}{c}{Baseline} & \multicolumn{2}{c}{Attenuation} & \multicolumn{2}{c}{Perfusion} & \multicolumn{2}{c}{Specific } & \multicolumn{2}{c}{Thermal} \\
          &		& \multicolumn{1}{c}{} & \multicolumn{2}{c}{} & \multicolumn{2}{c}{} & \multicolumn{2}{c}{heat capacity} & \multicolumn{2}{c}{conductivity} \\
          						&				&       & $-$SD   	& $+$SD & $-$SD & $+$SD   	& $-$SD & $+$SD   	& $-$SD & $+$SD \\
    \hline
	Maximum temperature 			& ($^{\circ}$C) 	& 72.3  	& 65.3		& 77.3  & 72.4  & 72.2  		& 74.5  & 70.3  		& 73.1  & 71.6 \\
    Difference (from baseline) 	& ($^{\circ}$C) 	& 0.0   	& $-$7.0  	& 5.0   & 0.1   & $-$0.1  	& 2.2   & $-$2.0  	& 0.8   & $-$0.8 \\
    Difference (from baseline) 	& (\%) 			& 0.0   	& $-$9.7  	& 6.9   & 0.2   & $-$0.2  	& 3.1   & $-$2.7  	& 1.1   & $-$1.0 \\
    \hline
    Thermal dose volume 			& (cm$^{3}$) 	& 2.6   	& 2.3   		& 2.7   & 2.7   & 2.5   		& 3.1   & 2.2   		& 2.6   & 2.5 \\
    Difference (from baseline) 	& (cm$^{3}$) 	& 0.0   	& $-$0.2  	& 0.1   & 0.1   & $-$0.1  	& 0.5   & $-$0.3  	& 0.1   & 0.0 \\
    Difference (from baseline) 	& (\%) 			& 0.0   	& $-$9.1  	& 5.8   & 4.5   & $-$2.9  	& 20.7  & $-$12.8 	& 2.1   & $-$1.6 \\
    \hline
    \end{tabular}
    }
  \label{tab:data}
\end{table*}

\begin{figure}[b!]
    \centering
    \subfigure[]
    {
        \includegraphics[width=0.44\columnwidth]{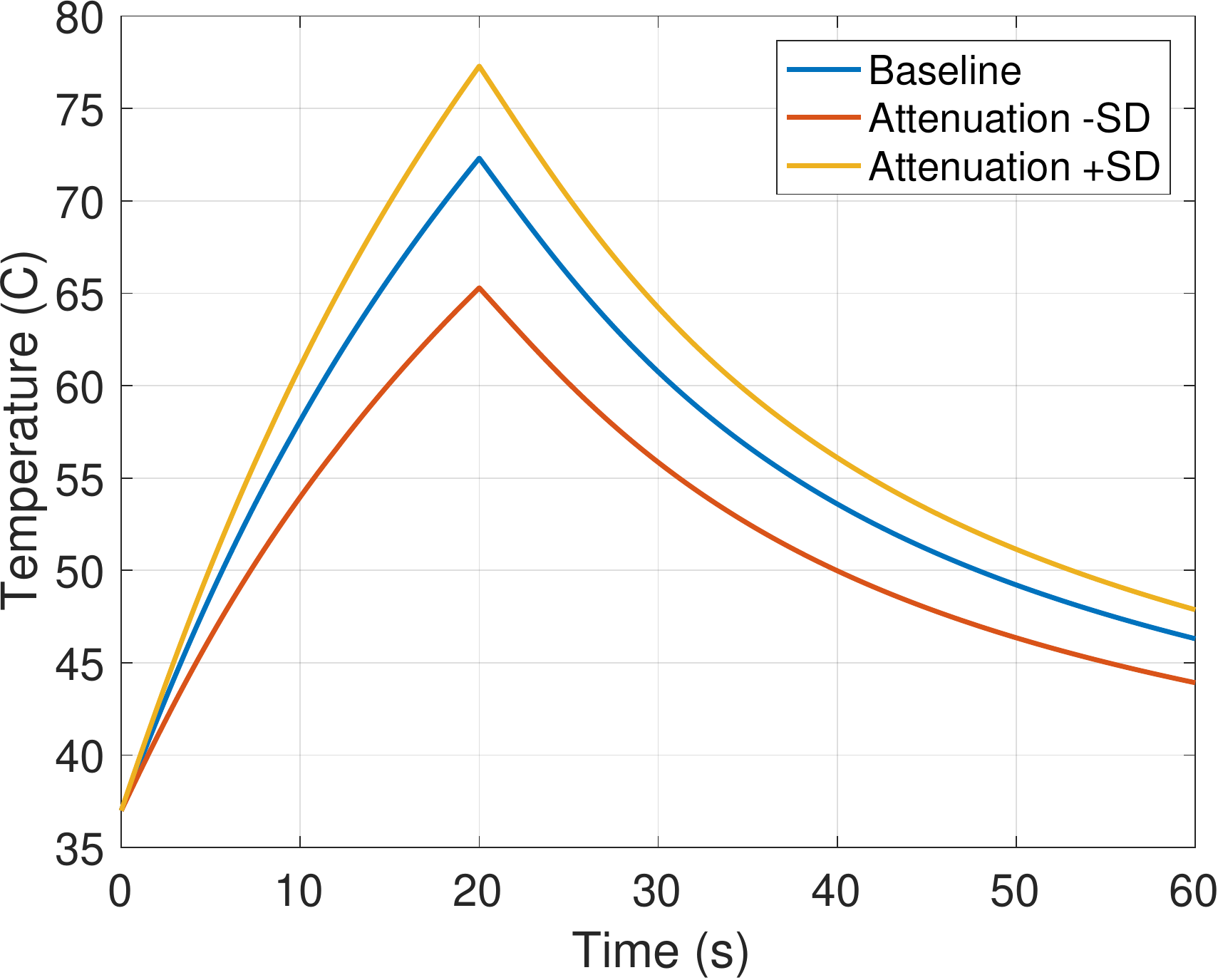}
    }
    \subfigure[]
    {
        \includegraphics[width=0.44\columnwidth]{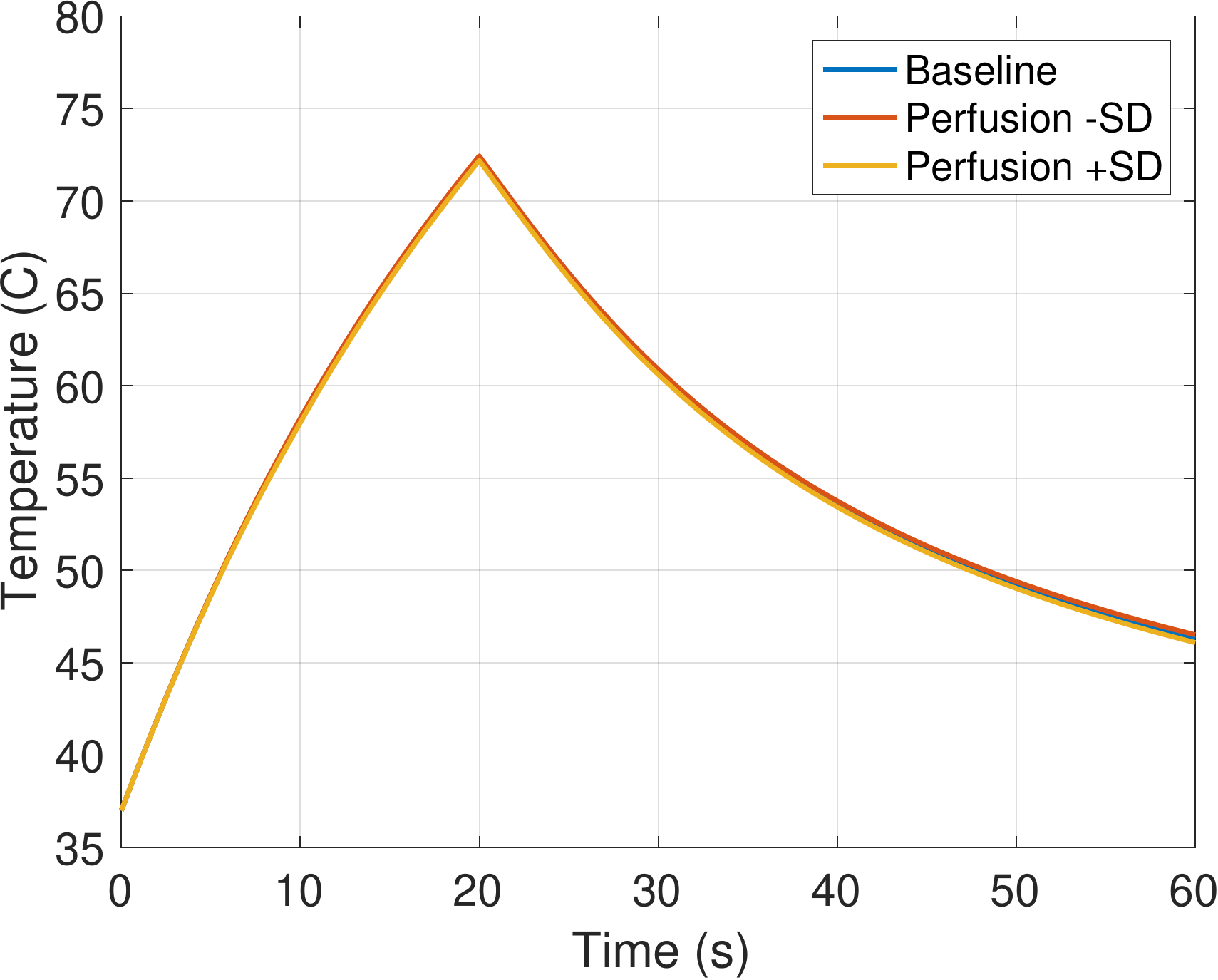}
    }
    \\
    \subfigure[]
    {
        \includegraphics[width=0.44\columnwidth]{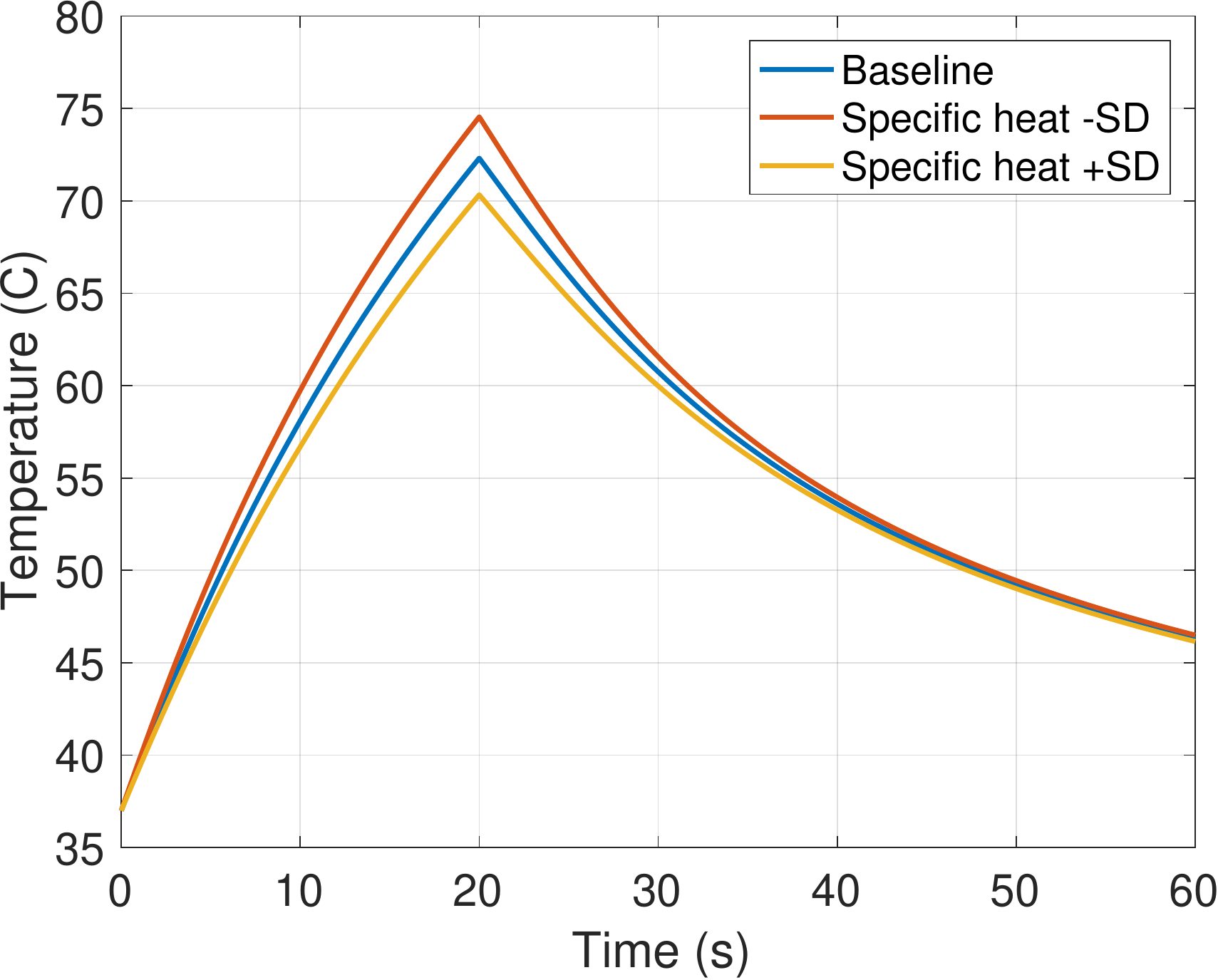}
    }
    \subfigure[]
    {
        \includegraphics[width=0.44\columnwidth]{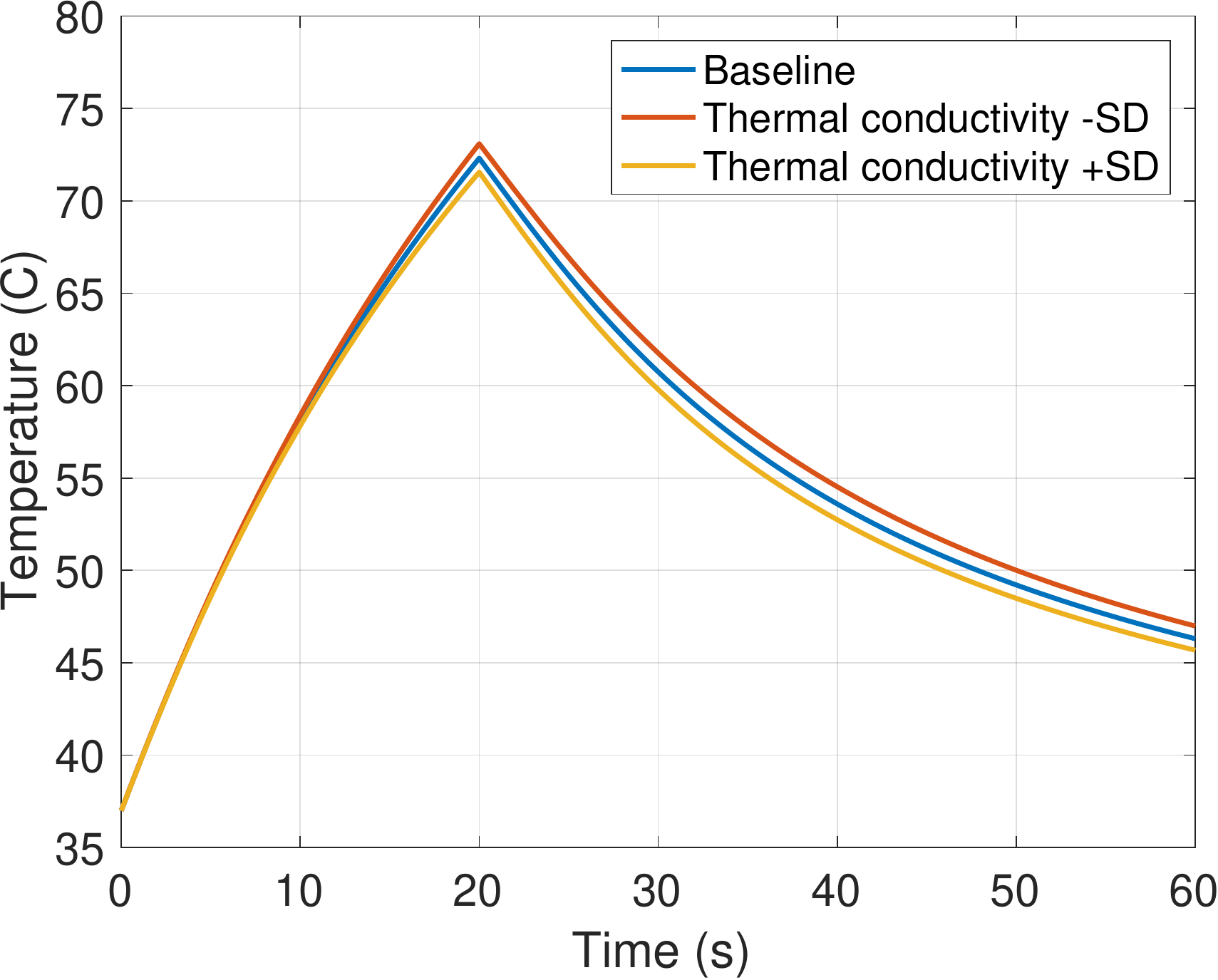}
    }
    \caption{The effect of the variation in (a) attenuation, (b) perfusion, (c) specific heat capacity and (d) thermal conductivity on temperature during 20-second sonication followed by a 40-second cooling period.}
    \label{fig:temp_time}
\end{figure}

\begin{figure}[b!]
    \centering
    \subfigure[]
    {
        \includegraphics[height=4cm]{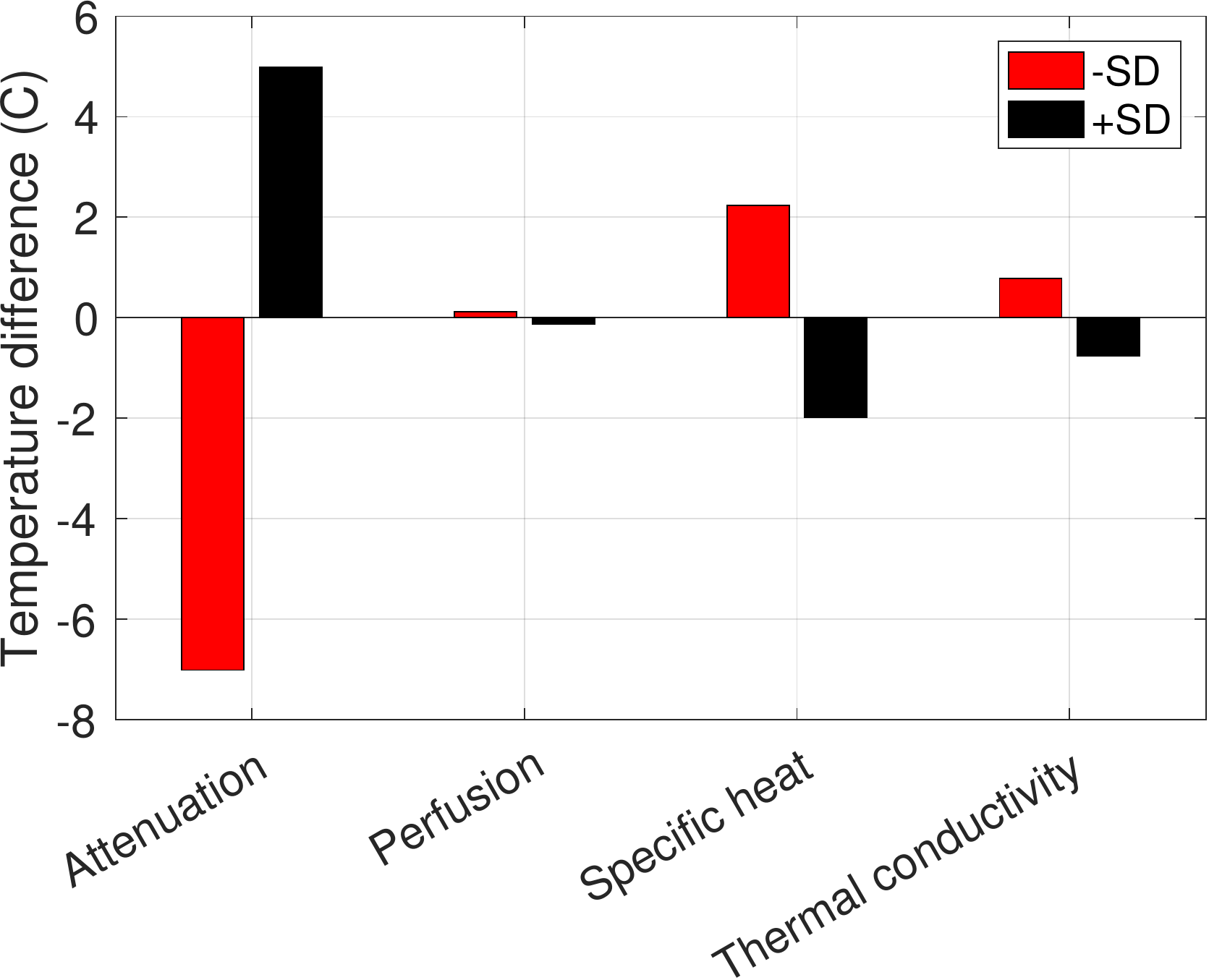}
    }
    \subfigure[]
    {
        \includegraphics[height=4cm]{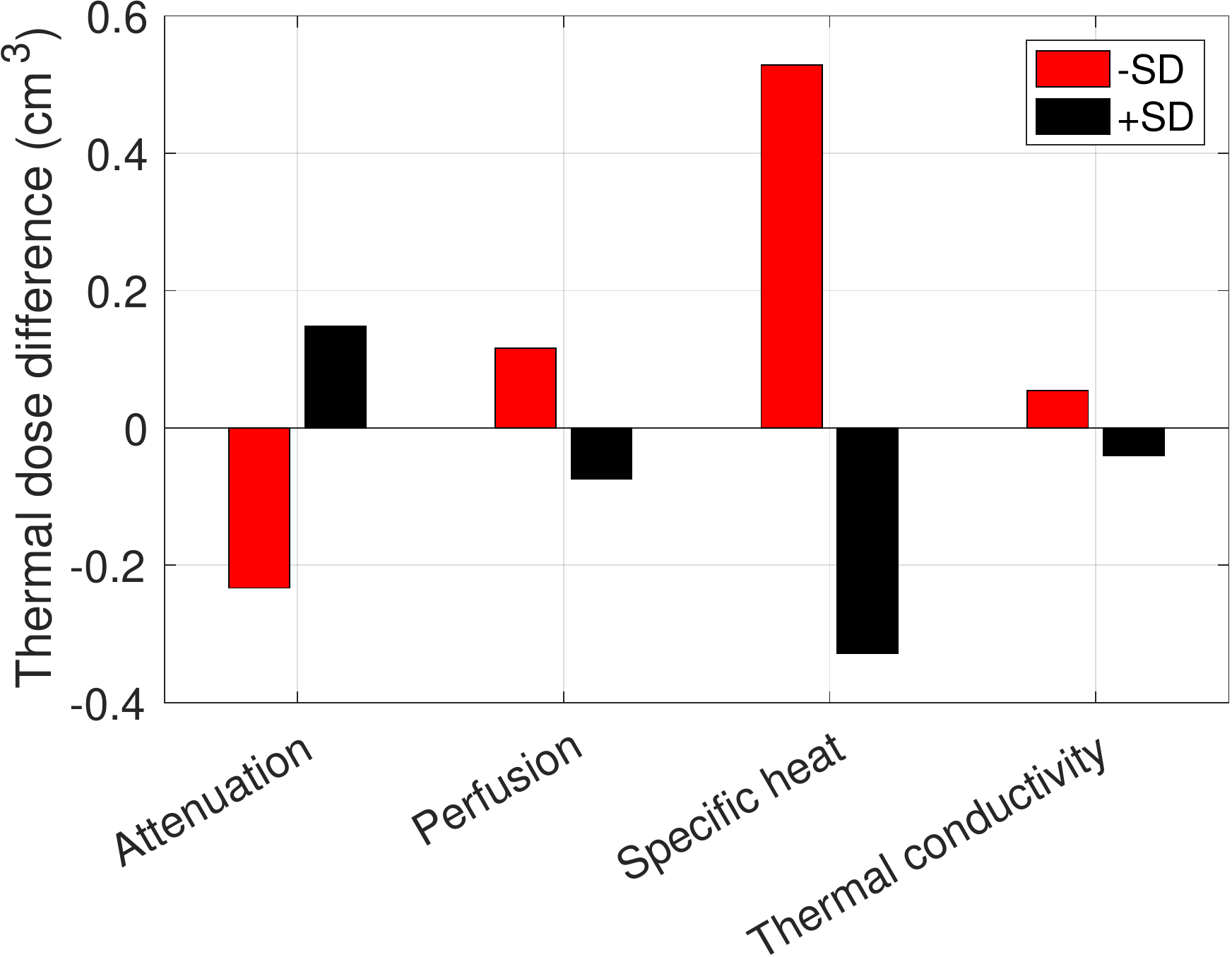}
    }
    \\
    \subfigure[]
    {
        \includegraphics[height=4cm]{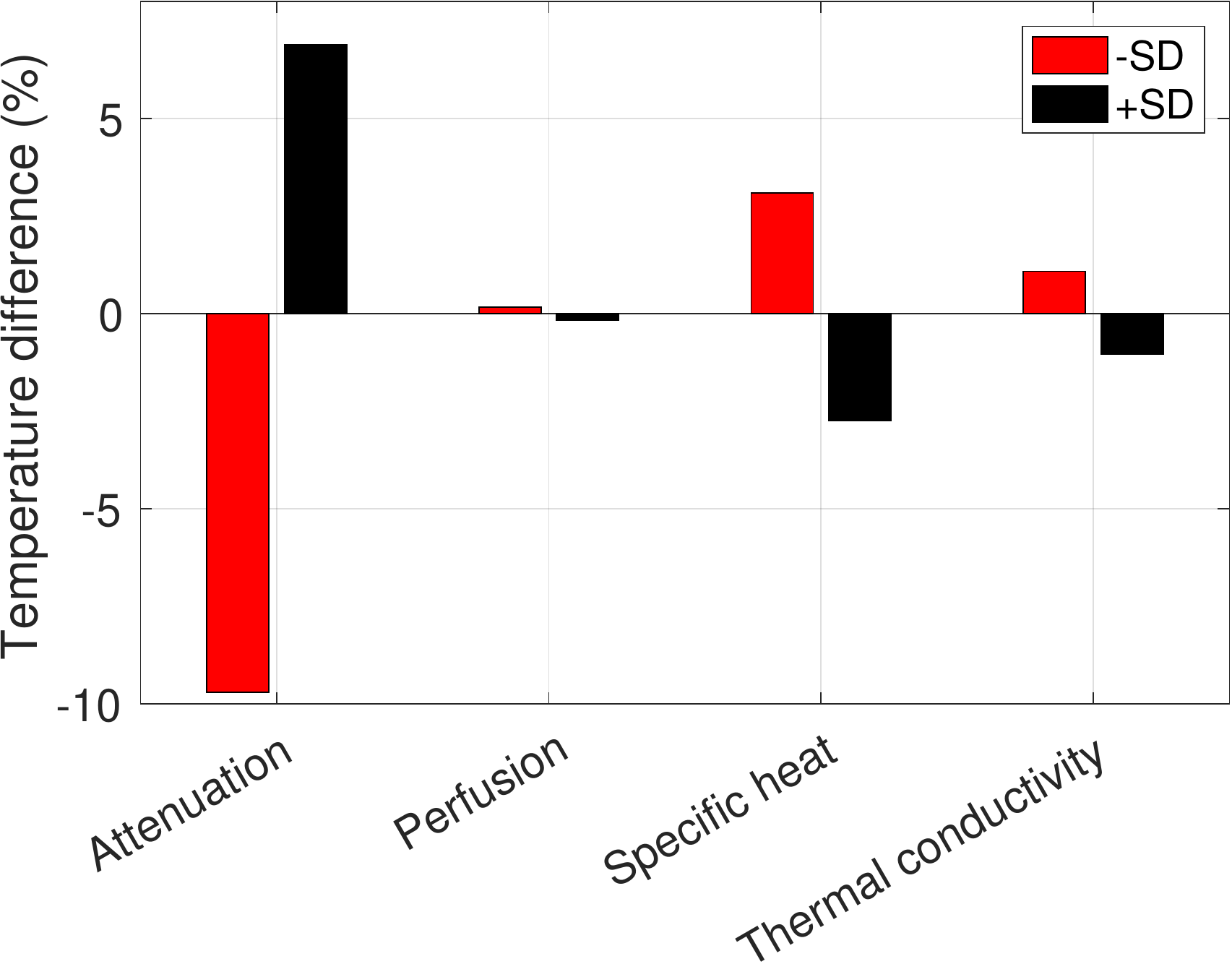}
    }
    \subfigure[]
    {
        \includegraphics[height=4cm]{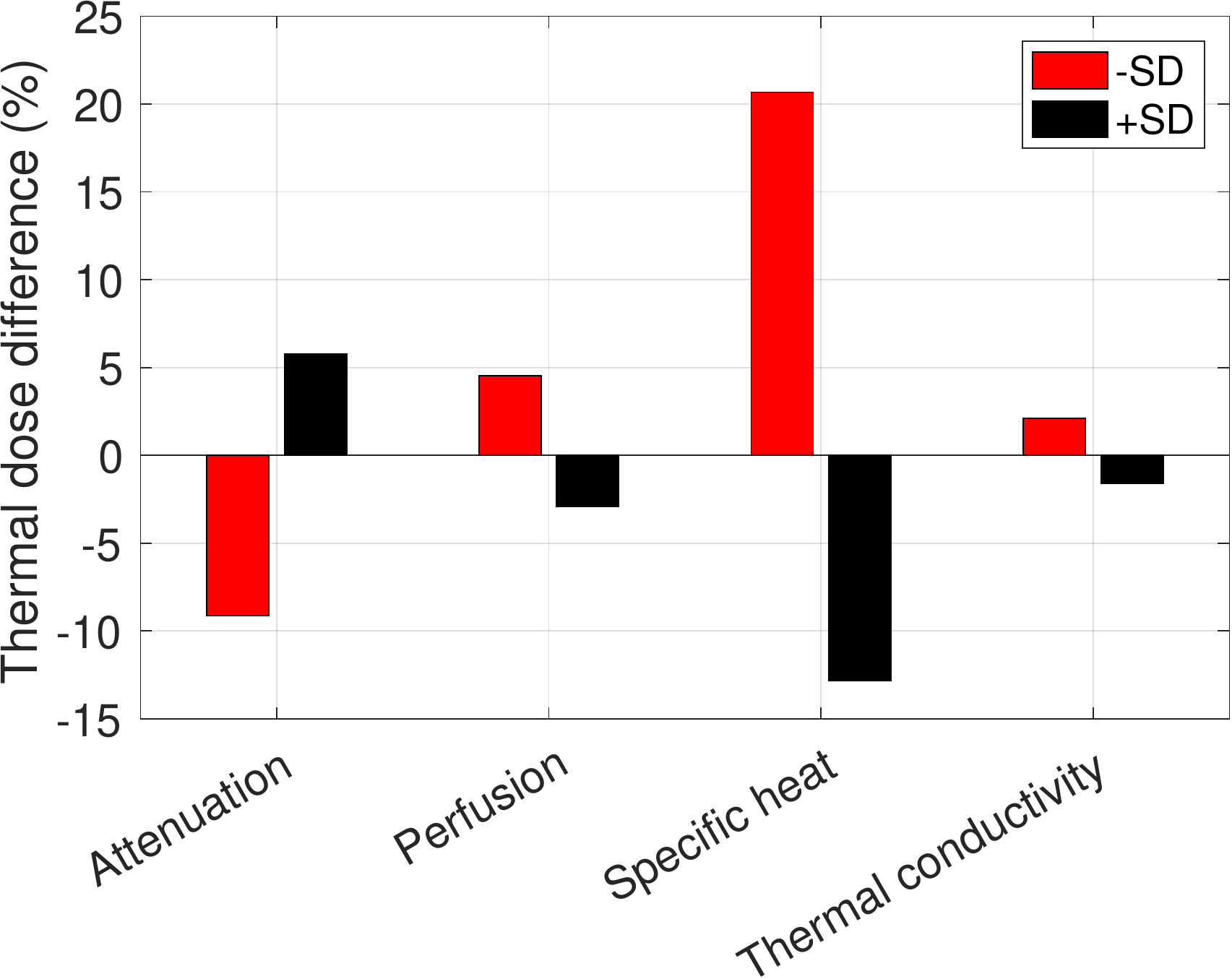}
    }
    \caption{The difference in (a) peak temperature and (b) thermal dose volume with respect to the baseline value when attenuation, perfusion, specific heat capacity and thermal conductivity are varied within their physical ranges. (c)-(d) The same graphs normalised to the baseline value.}
    \label{fig:bar_temp_cem}
\end{figure}

Figure \ref{fig:temp_time} shows the temperature evolution during the heating and cooling recorded at 5 mm away from the geometric centre of the transducer (the origin in the coordinates). The baseline curve in each figure corresponds to the simulation with mean values while the two other curves show the specific tissue property changed by $\pm$SD. In Figure \ref{fig:bar_temp_cem} are shown the differences in peak temperature and thermal dose with respect to the baseline sonication. Figures \ref{fig:bar_temp_cem}(a)-(b) show the absolute differences in temperature and thermal dose, respectively, while Figures \ref{fig:bar_temp_cem}(c)-(d) are normalised to the baseline value. Table \ref{tab:data} lists the corresponding numerical values for maximum temperature and thermal dose in each individual case.

Figure \ref{fig:temp_time}(a) shows the effect of attenuation on heating during the sonication. Increasing the attenuation by one SD resulted in 6.9\% higher peak temperature at the end of the sonication. This is because the heating rate of the ultrasound field is directly proportional to absorption. Similarly, decreasing the attenuation by one SD resulted in a 9.7\% decrease in the peak temperature. The corresponding 240~CEM thermal doses for increased and decreased attenuation exhibited similar behaviour with a 5.8\% and $-$9.1\% change from the baseline, respectively.

The effect of perfusion on temperature evolution is shown in Figure \ref{fig:temp_time}(b). For the given perfusion values, the effect on heating is negligible. Changing perfusion $\pm$SD from the baseline resulted in approximately $\mp$0.2\% deviation in maximum temperature. This is likely due to the fact that the perfusion was relatively small to begin with and that the perfusion diminished to zero quite fast in the regions where 240~CEM thermal dose was achieved. The corresponding effect on thermal dose was a magnitude larger with a $-$2.9\% and 4.5\% change for increased and decreased perfusion rate, respectively. This is due to the cooling effect of perfusion in the tissue regions which are surrounding the necrotic (i.e., 240 CEM) tissue thus reducing its growth speed.

In Figure \ref{fig:temp_time}(c) is shown the effect of varying specific heat capacity. Heat capacity specifies the amount of thermal energy needed to increase the temperature of the tissue. Therefore, the reduction in heat capacity results higher peak temperature and vice versa, with the corresponding changes in the maximum temperature being 3.1\% and $-$2.7\%, respectively. The effect on thermal dose was drastically larger with 20.7\% and $-$12.8\% deviations from the baseline for the decreased and increased heat capacity, respectively. This means that additional tissue regions were able to achieve sufficient temperatures to exceed the thermal dose threshold with the given sonication duration.

The last tissue parameter studied was thermal conductivity which is shown in Figure \ref{fig:temp_time}(d). Increasing thermal conductivity by one SD resulted in 1.0\% decrease in peak temperature. A similar effect was seen with decreased thermal conductivity which increased the peak temperature by 1.1\%. The effect on thermal dose was similar with a $-$1.6\% and 2.1\% change from the baseline with increased and decreased thermal conductivity, respectively.

\newpage

\section*{Discussion}

Among the studied tissue parameters, the biggest effect on peak temperature was caused by the changes in attenuation. One SD increase in attenuation resulted in a 6.9\% higher peak temperature. The temperature evolution was studied at 5~mm from the transducer face where the increase in attenuation shows as a higher heating rate. On the other hand, tissue regions further away from the transducer exhibit lower heating due to a reduction in intensity with ultrasound propagation distance. Therefore, higher attenuation of the prostate shows up as increased heating in locations close to the transducer.

Perhaps surprisingly, the changes in perfusion had the smallest effect on peak temperature. As mentioned earlier, the perfusion in the prostate is relatively small to start with (about one sixth of that in kidney medulla \citep{roberts1995renal}), and thus, the changes to it did not have a big effect on the temperature evolution. Furthermore, the perfusion was set to zero in the tissue regions over 240~CEM which further diminished its effect on temperature. These results are consistent with earlier studies which show that perfusion does not have noticeable effect on the predictability and efficacy of the ultrasound therapy in the prostate \citep{rouviere2004can, kirkham2008mr}.

The biggest effect on 240~CEM thermal dose was seen in the variations of specific heat capacity. One SD decrease in specific heat capacity showed 20.7\% increase in thermal dose volume. The smallest effect on thermal dose was caused by the changes in thermal conductivity. These values present the scale of variation might be expected during a clinical treatment.

It should be noted that some of the tissue properties are also temperature dependent, which the simulation model did not take into account \citep{van2002prostate}. Furthermore, tumorous tissue might exhibit different properties to healthy tissue \citep{inaba1992quantitative}.

\section*{Conclusions}

The effect of natural physiological variation in attenuation, perfusion, specific heat capacity and thermal conductivity on the efficacy of therapeutic ultrasound treatment in the prostate was studied. It was found that with the given sonication duration, attenuation had the biggest effect on temperature while perfusion had the smallest. Thermal dose was mostly affected by the variations in specific heat capacity whereas thermal conductivity had the smallest effect.

\section*{Acknowledgements}

V.~S. acknowledges the support of the State Research Funding (ERVA), Hospital District of Southwest Finland, number K3007 and CSC - IT Center for Science, Finland, for computational resources. J.~J. was supported by The Ministry of Education, Youth and Sports of the Czech Republic from the National Programme of Sustainability (NPU II); project IT4Innovations excellence in science - LQ1602. B.~T. acknowledges the support of EPSRC grant numbers EP/L020262/1 and EP/M011119/1.

\pagebreak

\bibliographystyle{UMB-elsarticle-harvbib}


\end{document}